\documentclass[a4paper]{article}

\usepackage{geometry}                
\usepackage{amsmath}                 
\usepackage{amssymb}                 
\usepackage{amsbsy}
\usepackage{amsthm}
\usepackage{amsfonts}

\usepackage{graphicx}
\usepackage{subfigure}
\usepackage{epstopdf,cite}

\usepackage[colorlinks,linkcolor=red]{hyperref} 

\usepackage{orcidlink}					

\usepackage{comment}

\usepackage{caption}		
\usepackage{makecell}		

\begin{document}
	
\title{\textbf{\emph{Velde}: constructing cell potential landscapes by RNA velocity vector field decomposition}}

\author{Junbo Jia$^{1}$\orcidlink{0000-0002-8983-2676}, Luonan Chen$^{1,2}$\thanks{Corresponding author. Email address: lnchen@sibcb.ac.cn} \\ \\
	$^1${\small Key Laboratory of Systems Health Science of Zhejiang Province, School of Life Science,}\\
		{\small Hangzhou Institute for Advanced Study, University of Chinese Academy of Sciences,}\\ 
		{\small Hangzhou 310024, China}\\
	$^2${\small Center for Excellence in Molecular Cell Science, Shanghai Institute of Biochemistry and Cell Biology,}\\
		{\small Chinese Academy of Sciences, Shanghai 200031, China}\\
}

\date{(\today)}


\maketitle

\begin{abstract}
	
	The Waddington landscape serves as a metaphor illustrating the developmental process of cells, 
	likening it to a small ball rolling down various trajectories into valleys. Constructing an 
	epigenetic landscape of this nature aids in visualizing and gaining insights into cell differentiation. 
	Development encompasses intricate processes involving both cell differentiation and cell cycles. 
	However, current landscape methods solely focus on constructing a potential landscape for cell differentiation, 
	neglecting the accompanying cell cycle. This paper introduces a novel method that simultaneously constructs 
	two types of potential landscapes using single-cell RNA sequencing data. Specifically, it presents the 
	natural Helmholtz-Hodge decomposition (nHHD) of a continuous vector field within a bounded domain in 
	$n-$dimensional Euclidean space. This decomposition uniquely breaks down the vector field into a gradient 
	field, a rotation field, and a harmonic field. Utilizing this approach, the RNA velocity vector field is 
	separated into a curl-free component representing cell differentiation and a curl component representing 
	the cell cycle. By calculating the corresponding potential functions, potential landscapes for both cell 
	differentiation and the cell cycle are obtained. Finally, the efficacy of this method is demonstrated 
	through its application to synthetic and real datasets.\\

\end{abstract}


\section{Introduction}
\label{sec-1}

\begin{figure}[!ht]
	\centering
	\includegraphics[width=0.9\linewidth]{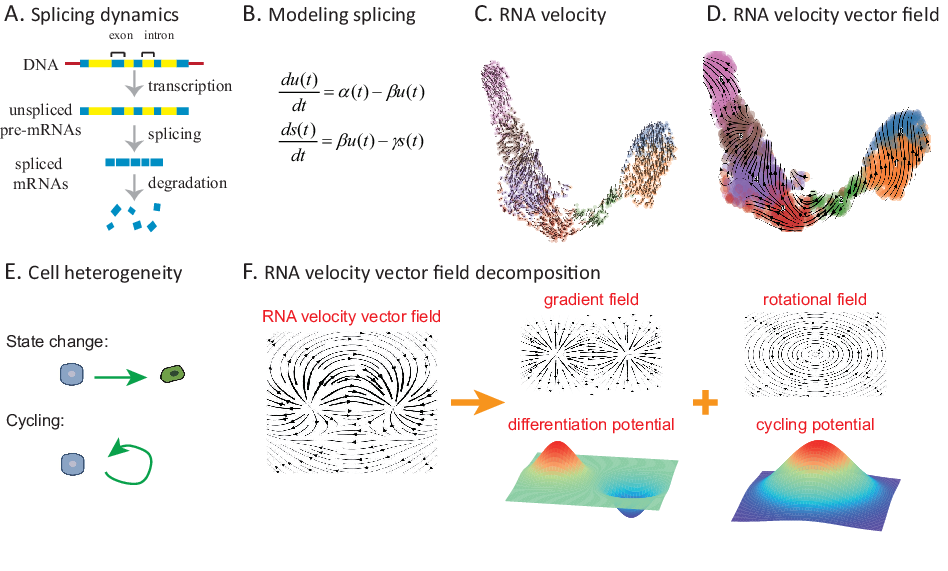}
	\caption{Schematic diagram of RNA-velocity and RNA-velocity vector field decomposition.}
	\label{fig-1}
\end{figure}

The invention of single-cell RNA sequencing (scRNA-seq) has enabled the study of cell heterogeneity in 
tissues and the development of cells at the resolution of individual cells \cite{tang2009mrna, eberwine2014promise}.
Particularly in medical research, scRNA-seq allows for more precise targeting of immunotherapy or targeted 
therapy towards specific cells or cell subpopulations \cite{rood2022impact,Satyen2020Applying,Lim:2023aa}. 
Consequently, scRNA-seq holds significant potential for broader application in analyzing biological functional 
mechanisms and discovering more effective methods for disease treatment.


scRNA-seq captures a snapshot of cell development, enabling the inference of developmental trajectories 
as one of the primary downstream analysis tasks (the other being cell subpopulation identification through 
clustering) \cite{luecken2019current}. By utilizing first principles, dynamic modeling of mRNA splicing 
during gene expression can yield RNA velocity, allowing for the determination of each cell's developmental 
direction in the near future \cite{la2018rna}, as depicted in Figure \ref{fig-1}ABC. Furthermore, leveraging
reconstruction methods for continuous vector fields, such as sparse vector field consuses (sparse VFC) 
\cite{ma2013regularized}, enables the construction of a comprehensive RNA velocity vector field, as 
illustrated in Figure \ref{fig-1}D, offering an analytical approach to studying cell state transitions 
\cite{qiu2022mapping}.


The Waddington landscape, proposed in 1957, offers an elementary and intuitive description of cell development
\cite{waddington1957strategy}. It draws an analogy between the development of cells and a small ball rolling 
along various tracks from a mountain slope to a valley. This analogy vividly portrays cell differentiation as 
the process of small balls selecting different tracks. In fact, this landscape assigns each cell a differentiation 
potential energy, where cells at higher elevations possess greater differentiation potential energy compared 
to cells at lower altitudes. This suggests that cells with higher potential have the ability to differentiate 
into cells with lower potential energy \cite{teschendorff2021statistical}. Currently, there exist several methods 
for quantifying cell potency and inferring the differentiation trajectory of cells from scRNA-seq data. 
Some graph-based methods include Wishbone \cite{setty2016wishbone}, Diffusion pseudotime \cite{haghverdi2016diffusion},
Monocle/Monocle2 \cite{trapnell2014dynamics, qiu2017reversed} scEpath \cite{jin2018scepath}, and other. 
Additionally, there are entropy-based methods such as StemID \cite{grun2016novo}, SLICE \cite{guo2017slice}, 
SCENT \cite{teschendorff2017single}, and Markov chain entropy \cite{shi2020quantifying}. Moreover, dynamic 
modeling-based methods like PBA \cite{tusi2018population} and LDD \cite{shi2019quantifying} are also available.


Although numerous methods exist for quantifying cell potency, constructing a Waddington-like landscape to 
illustrate cell development still poses a challenge. Some approaches aim to build such landscapes based on gene 
regulatory networks (GRNs) \cite{wang2010potential, guo2017netland}, but they are constrained by the 
availability of prior information on gene regulation and the limited number of genes considered. 
Moreover, these methods are not applicable to real scRNA-seq data. In this article, we present a novel method 
that addresses these limitations by simultaneously constructing two landscapes: the cell differentiation 
landscape and the cell cycle landscape, using scRNA-seq data. We posit 
that the heterogeneity of cells primarily arises from two concurrent processes: transition and cycle, 
as illustrated in Figure \ref{fig-1}E. Transition can be described as a cell changing from one state to 
another, such as during differentiation and dedifferentiation. Meanwhile, cycle is described as the 
cell rotating around its state. Building on this assumption, we extend the natural Helmholtz-Hodge 
decomposition method in $2-$ and $3-$dimensional space \cite{bhatia2014natural} to $n-$dimensional space, 
and employ this method to decompose the RNA velocity vector field into the sum of a gradient vector field 
and a rotating vector field, which describe cell transition and cycle, respectively, in order to derive 
their differentiation potential landscape and cell cycle potential landscape, as depicted in Figure
\ref{fig-1}F. This method is termed \emph{Velde}, signifying its principle is to construct 
distinct potential landscapes through RNA \emph{Vel}ocity vector field \emph{de}composition. Finally, 
we validate the effectiveness of this method using both synthetic and real datasets.

The rest of the paper is organized as follows. In section \ref{sec-2}, the underlying theory and constructing 
process of landscapes was provided. In section \ref{sec-3}, the effectiveness of our method was verified using
different datasets. Finally, conclusions and discussions was given in section \ref{sec-4}.

\section{Methods}
\label{sec-2}

Before delving into the specific process of \emph{Velde}, let us first introduce the underlying theory.

\subsection{The natural Helmholtz-Hodge decomposition of $n-$dimensional vector field}
\label{sec-2-1}


Here we first propose the natural Helmholtz-Hodge decomposition (nHHD) of vector fields in an 
$n-$dimensional bounded domain, which is the cornerstone of our method \emph{Velde}. The nHHD is based 
on two works. First, Gl{\"o}tzl Erhard and Richters Oliver considered the rotation as a superposition 
of $\binom{n}{2}$ rotations within the coordinate planes and extend the Helmholtz-Hodege decomposition 
to vector fields in $n-$dimensional Euclidean space \cite{glotzl2020helmholtz}. 
Second, Bhatia Harsh et al. proposed original nHHD in 2- or 3-dimensional space by separating the flows 
into internal and external components, which provides a unique and reliable decomposition while avoiding 
boundary condition constraints \cite{bhatia2014natural}. 

The nHHD of vector field in the $n-$dimensional bounded domain is described as: for any bounded domain 
$\Omega\subset \mathbb{R}^n$ and any twice continuously differentiable vector field 
$f: \Omega \rightarrow \mathbb{R}^n$ that decays faster than $\| x \|^{-c}$ for 
$\|x\|\rightarrow \infty$, due to internal and external influences, it is decomposed into three vector fields:
\begin{equation}\label{eq-1}
	f(x)=g(x) + r(x) + h(x).
\end{equation}
Where, $g(x)$ is a gradient field, where its flow emits out of the source and into the sink, and it is 
irrotational. $r(x)$ is a rotational field, which tends to rotate locally. And $h(x)$ is a harmonic field 
that has the properties both of gradient and rotational fields.


$g(x)$ and $r(x)$ are considered as natural gradient field and natural rotational field, respectively, 
representing the fields that are within the domain $\Omega$ and influenced by the source and rotation 
within this domain. However, $h(x)$ is a harmonic field, representing the field that is within the domain 
$\Omega$ but influenced by the exterior.


For a given domain $\Omega\subset \mathbb{R}^n$ and vector field $f$ defined on it, 
the calculation of nHHD can be divided into the following three phases, as shown in Figure \ref{fig-2}A:

\begin{enumerate}
	\item \textbf{From vector field to densities.} Calculate the scalar source density $\gamma(x)$ and rotation 
	density matrix $\rho(x)$ in the domain $\Omega$:
	\begin{subequations}
		\begin{align}
			\gamma(x) &= \mathrm{div} f(x) \label{eq-2a} \\
			\rho(x)	  &= \overline{\mathrm{ROT}} f(x) = [\rho_{ij}] \label{eq-2b}
		\end{align}
	\end{subequations}

	\item \textbf{From densities to potentials.} The convolution of the densities with the fundamental solution 
	of the Laplace equation in domain $\Omega$ provides the natural source potential $G(x)$ and the natural rotation 
	potential matrix $R(x)$:
	\begin{subequations}
		\begin{align}
			G(x) 	&= \overline{\int}_{\Omega} \gamma(x) = \int_\Omega \Gamma(x-y)\gamma(y)\mathrm{d}\omega(y) \label{eq-3a}\\ 
			R(x)	&= \bigg[\overline{\int}_{\Omega} \rho_{ij}(x)\bigg] = \bigg[\int_\Omega \Gamma(x-y)\rho_{ij}(y)\mathrm{d}\omega(y)\bigg]	\label{eq-3b}
		\end{align}
	\end{subequations}
			
	\item \textbf{From potentials to vector fields.} By calculating the gradient and rotation of the 
	corresponding potentials, the gradient field $g(x)$ and rotation field $r(x)$ can be obtained, 
	respectively, and the remaining fields of $f(x)$ are harmonic fields $h(x)$:
\end{enumerate}
	\begin{subequations}
		\begin{align}
			g(x) 	&= \mathrm{grad} G(x) \label{eq-4a}\\ 
			r(x)	&= \mathrm{ROT} R(x) \label{eq-4b}\\ 
			h(x)	&= f(x)-g(x)-r(x) \label{eq-4c}
		\end{align}
	\end{subequations}

Here, the related symbols and their meanings are shown in Supplementary Table \ref{stab-1}. 
In addition, according to the above calculation phases and formulas, the densities are uniquely determined by domain 
$\Omega$ and vector field $f$, so that the potentials and their subsequent fields are also uniquely determined. 
Therefore, nHHD is unique.

\subsection{Constructing cell differentiation potential and cycle potential by \emph{Velde}}

\begin{figure}[!ht]
	\centering
	\includegraphics[width=0.9\linewidth]{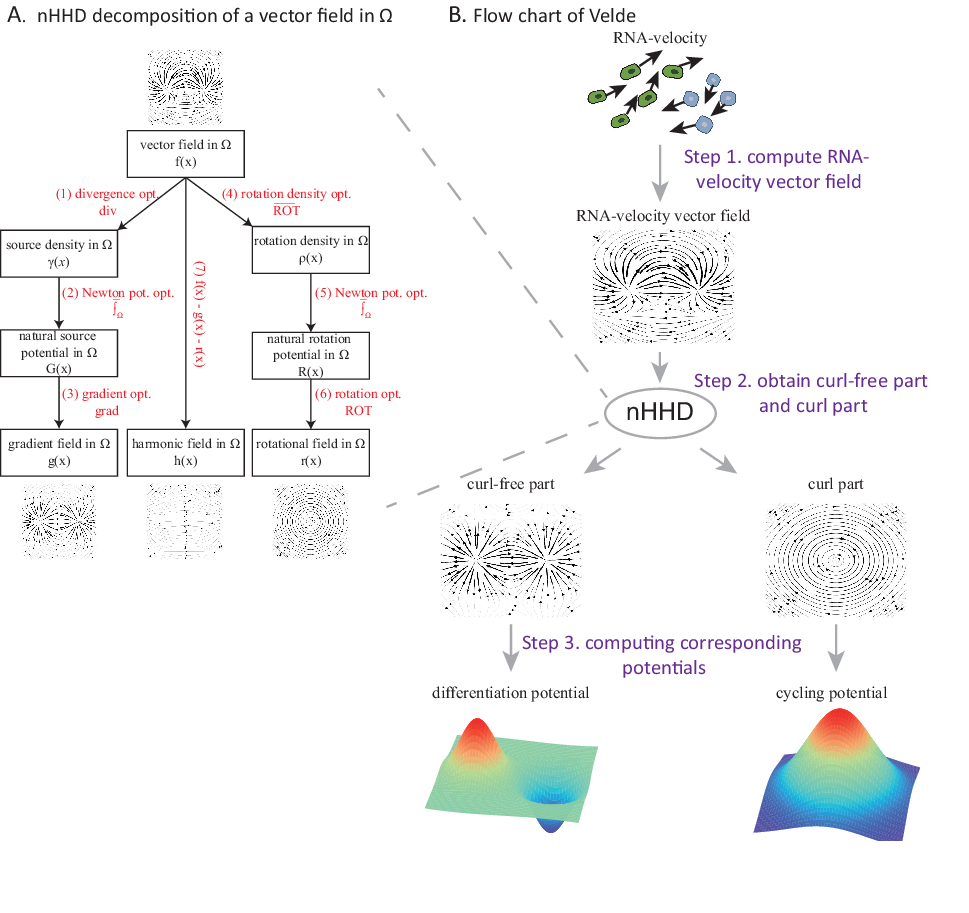}
	\caption{Natural Helmholtz-Hodge decomposition and flow chart of \emph{Velde}.}
	\label{fig-2}
\end{figure}

%
%
%
%
%


Previous studies on potential landscape only attempted to construct one landscape, which can been considered 
as potential landscape of cell differentiation. By contrast, here we will construct two potential landscapes, a cell differentiation 
potential landscape and a cycle potential landscape. This is mainly based on the fact that if we do 
not consider random factors such as external environmental disturbance and noise in expression dynamics, the 
heterogeneity of cells in tissues mainly comes from cell differentiation and cell cycle. 
Therefore, it is very appropriate to establish potential landscapes that correspond to these two processes 
separately.


\emph{Velde} construct differentiation potential landscape and cycle potential landscape by decomposing the 
RNA-velocity vector field.  Its  input is RNA-velocity from scRNA-seq data, which can be obtained through 
third party libraries, such as \emph{velocyto} \cite{la2018rna} and  \emph{scVelo} \cite{bergen2020generalizing}. 
The results are differentiation potential landscape and cycle potential landscape. As shown in Figure \ref{fig-2}B, 
the calculation process consists of three steps:


\begin{description}
	\item[Step 1:] \textbf{Calculate the RNA-velocity vector field in the low-dimensional embedding space.}
	In theory, the low-dimensional embedding space can be any embedding, but embeddings that can maintain 
	the neighborhood relationship in the high-dimensional gene space are recommended, such as Uniform 
	Manifold Approximation and Projection (UMAP) and t-distributed Stochastic Neighbor Embedding (t-SNE). 
	For subsequent calculation, we use the velocity on the grid to discretize the RNA-velocity vector field, 
	which can be obtained using the Gaussian kernel method, or firstly estimating the analytical expression of 
	the vector field by sparse VFC and then calculating the vector field on the grid.
	
	\item[Step 2:] \textbf{Obtain the curl-free part and curl part of the RNA-velocity vector field.}
	By the nHHD method proposed in the previous subsection, the velocity vector field is uniquely 
	decomposed into three vector fields: gradient field $g(x)$, harmonic field $h(x)$, and rotation field $r(x)$. 
	Since the first two fields can be explained as the cell transformation towards a direction away from its 
	own current state, we merged these two fields to form a curl-free part to describe the cell differentation. 
	And the remaining rotation field serves as the curl part, which describe the cell cycle.
	
	\item[Step 3:] \textbf{Calculate the differentiation potential and cycle potential.}
	Due to the fact that the curl-free part is a superposition of a gradient field and a harmonic field, 
	and it is still a gradient field. Therefore, theoretically, it has a corresponding source potential 
	function, which we consider as the cell differentiation potential and it can be obtained by integrating 
	the curf-free part. The rotational potential corresponding to the curl part serves as the cycle potential, 
	which can be directly assigned to the natural rotational potential in the previous nHHD decomposition.
\end{description}

Note that the differentiation potential landscape here is the same as the usual potential landscape, 
describing the differentiation of cells from high potential areas to low potential areas, along the 
negative gradient of the landscape, $-\mathrm{grad} G(x)$, and the greater the potential difference, 
the easier it is for cells to differentiate. The cell cycle potential, nevertheless, is different in that 
it describes the cell cycle along the direction of rotation operator, $\mathrm{ROT} R(x)$, that is, along the 
clockwise contour of the cycle potential. And the steeper the cycle potential with cone-shape, 
the stronger the cell cycle.

\section{Results}
\label{sec-3}

To verify the effectiveness, \emph{Velde} was applied to both synthetic and real datasets.

\subsection{Identifying hidden components of a synthetic vector field}

\begin{figure}[!ht]
	\centering
	\includegraphics[width=0.9\linewidth]{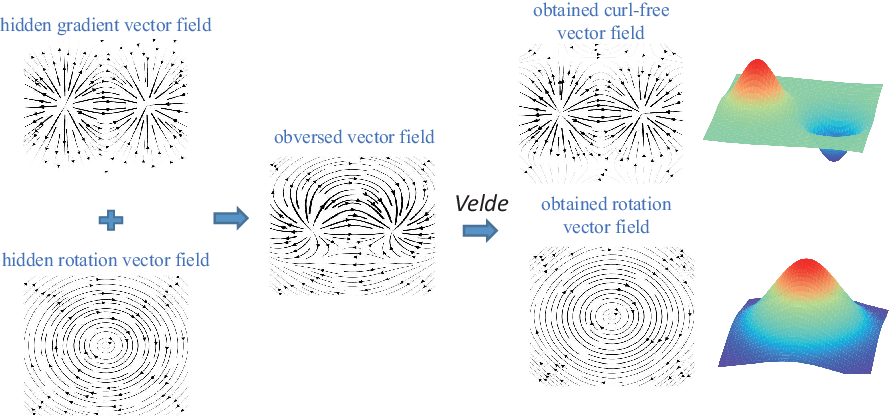}
	\caption{A synthetic vector field.}
	\label{fig-3}
\end{figure}

In this subsection, we considered two synthetic datasets.
The first is constructed by combining a hidden gradient vector field and a hidden rotation vector field, 
as shown in Figure~\ref{fig-3}. Applying \emph{Velde} to this dataset yields an curl-free vector field and
a curl vector field. The curf-free part  is almost identical to the hidden gradient vector field except for some slight 
differences in the boundary area, and the obtained  rotation part is also very similar to the hidden rotation 
vector field as well. This indicates that \emph{Velde} can indeed effectively extract the hidden gradient and rotational 
components of the vector field.

\begin{figure}[!ht]
	\centering
	\includegraphics[width=0.9\linewidth]{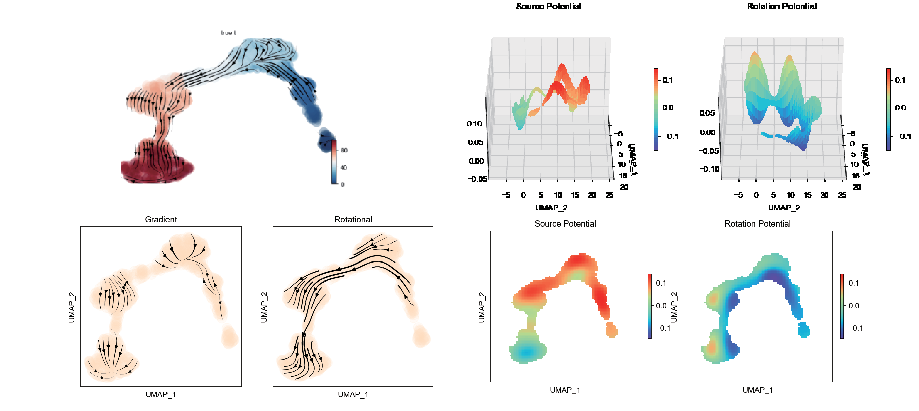}
	\caption{A synthetic scRNA-seq dataset.}
	\label{fig-4}
\end{figure}

The second is a synthetic scRNA-seq dataset, with the differentiation direction of cells shown in the RNA velocity 
in Figure~\ref{fig-4}. By applying \emph{Velde}, two potential landscapes were obtained, with the source potential on the left 
and the rotational potential on the right corresponding to the cell differentiation potential and cycle potential, 
respectively. From the differentiation potential landscape, it can be seen that the cells on the right generally have 
higher potential than the cells on the left, which is consistent with their differentiation direction. 
In addition, no cell regions resembling conical peaks were found within the cell population in the cycle potential 
landscape, indicating no obvious cell cycle.

\subsection{Constructing cell differentiation potential and cycle potential from real datasets}

\begin{figure}[!ht]
	\centering
	\includegraphics[width=0.9\linewidth]{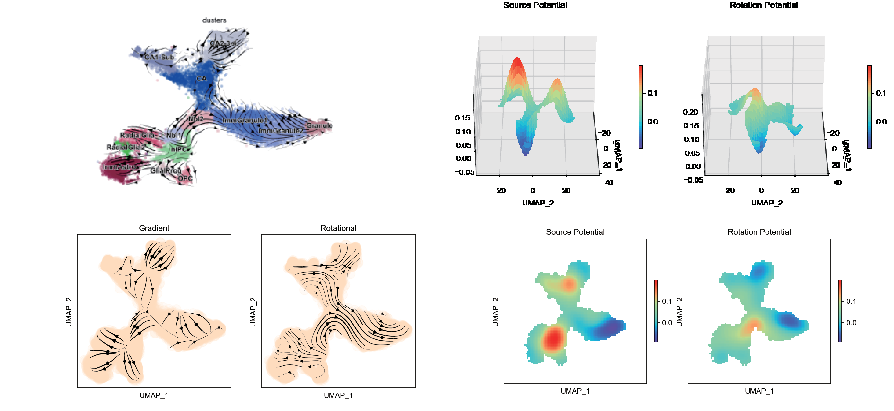}
	\caption{Dentategyrus scRNA-seq dataset~\cite{la2018rna}.}
	\label{fig-5}
\end{figure}

\begin{figure}[!ht]
	\centering
	\includegraphics[width=0.9\linewidth]{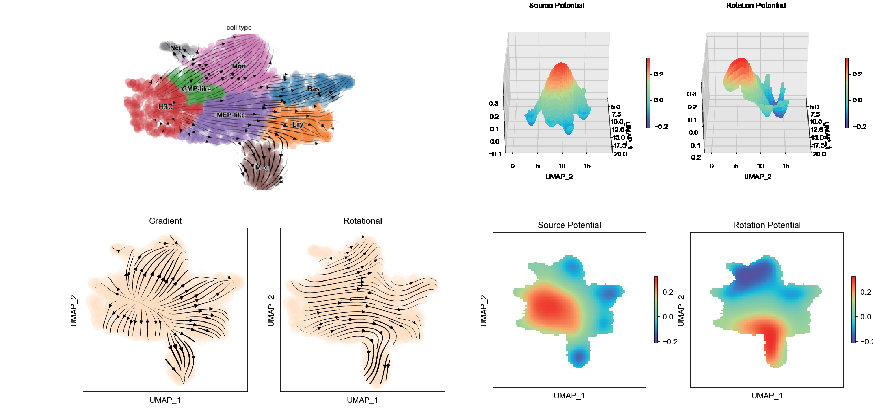}
	\caption{Human hematopoiesis scRNA-seq dataset~\cite{qiu2022mapping}.}
	\label{fig-6}
\end{figure}

We also applied \emph{Velde} to three real scRNA-seq dataset, with the first two datasets coming from 
works related dentategyrus lamanno~\cite{la2018rna} and human hematopoiesis~\cite{qiu2022mapping}, respectively.
The developmental directions and trajectories of cells are shown in the RNA velocity in the Figure \ref{fig-5} and \ref{fig-6}. 
The differentiation potential obtained by \emph{Velde} is consistent with the actual differentiation path of cells, 
and the obtained cell cycle potential also indicates that both have no obvious cell cycle.

\begin{figure}[!ht]
	\centering
	\includegraphics[width=0.9\linewidth]{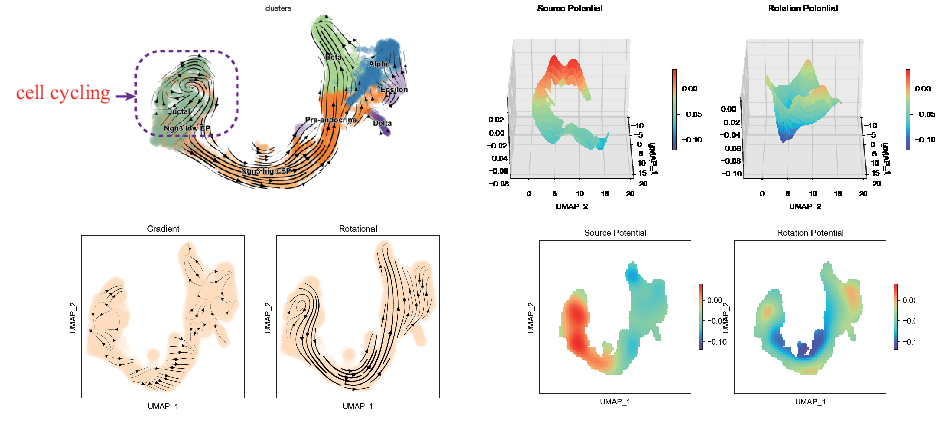}
	\caption{Pancreas scRNA-seq dataset~\cite{bastidas2019comprehensive}.}
	\label{fig-7}
\end{figure}

The third real dataset is about pancreatic development~ \cite{bastidas2019comprehensive}. 
Unlike the first two real datasets, this dataset not only has cell differentiation but also has strong cell cycle, 
as shown in Figure~\ref{fig-7}. \emph{Velde} calculated the correct relative differentiation potential, 
correctly identified the cell cycle at the location of the left cells, and obtained a conical peak-shaped cell 
cycle potential landscape at the corresponding site.

\section{Conclusions and discussions}
\label{sec-4}

Constructing a reasonable potential landscape helps to display and understand tissue development. 
However, the current landscape methods only construct a cell developmental potential landscape, 
without considering the cell cycle associated with the cell differentiation during the development. 
In this article, the natural Helmholtz-Hodge decomposition of continuous vector fields on bounded domains 
in $n-$dimensional Euclidean space was proposed and a method based on this to simultaneously construct 
cell differentiation potential landscape and cycle potential landscape, called \emph{Velde}, was developed as well. 
The effectiveness of \emph{Velde} in calculating cell differentiation potential, identifying cell cycle, and constructing 
cell differentiation potential landscape and cell cycle potential landscape was verified through its application to 
synthetic and real datasets.

Although \emph{Velde} can construct both cell differentiation potential and cycle potential simultaneously, 
its accuracy is still influenced by multiple factors. The first one is RNA velocity. Due to various noises and 
technical limitations during the sequencing process, there are inevitably deviations or errors in the 
estimating of cell RNA velocity, which directly affects the accuracy of \emph{Velde}. The other is embedding. 
The reasonable layout of data points in the embedding space and whether the embedding captures the topology 
of cell differentiation in high-dimensional space can also affect the calculation and aesthetics of potential landscape. 
In addition, constructing the potential landscape is not the ultimate goal of analyzing scRNA-seq data. 
Based on potential landscapes, extracting the pathways involved, as well as the regulatory mechanisms between each 
pathway and gene is our research direction in the future.

%
%
%
%
%
\section*{Acknowledgements}
\label{sec-5}

This work was supported by the NSFC grants under Grant Nos. 12201150. We would like to thank
Tiejun Li for his helpful advice and discussions.


\section*{Conflict of interest}

All authors declare no conflicts of interest in this paper.

\bibliographystyle{apalike}

\bibliography{references}

\newpage
\appendix
\section{Supplementary information}

\subsection{Supplementary Table 1: symbols and meanings}

\begin{table}[!ht]
	\centering
	\caption{Some variables and operators along with corresponding explanations}\label{stab-1}
	\begin{tabular}{l|l}
		\hline
		\textbf{Variable or Operator} & \textbf{Explanation}  \\
		\hline\hline
		\makecell[l]{vector filed\\ $f\in C^2(\mathbb{R}^n, \mathbb{R}^n)$} &  $f(x)=\Big( f_k(x); 1\leq k \leq n\Big)$\\
		\hline
		\makecell[l]{divergence operator\\ $\mathrm{div}: C^1(\mathbb{R}^n, \mathbb{R}^n)\rightarrow C^0(\mathbb{R}^n, \mathbb{R})$} &  $ \mathrm{div}f(x)=\sum_{i=1}^{n}\frac{\partial f_i(x)}{\partial x_i}$ \\
		\hline
		\makecell[l]{scalar source density\\ $\gamma \in C^1(\mathbb{R}^n, \mathbb{R})$} & $\gamma(x)=\mathrm{div}(x)$  \\
		\hline
		\makecell[l]{Newton potential operator\\ $\overline{\int}: C^0(\mathbb{R}^n, \mathbb{R})\rightarrow C^1(\mathbb{R}^n, \mathbb{R})$} &  $\overline{\int}\gamma(x):=\Gamma(x)*\gamma(x)=\int_{\mathbb{R}^n}\Gamma(x-y)\gamma(y)\mathrm{d}\omega(y)$\\
		\hline
		\makecell[l]{scalar source potential\\ $G\in C^1(\mathbb{R}^n, \mathbb{R})$} &  $G(x)=\overline{\int}\gamma(x)$\\
		\hline
		\makecell[l]{gradient operator\\ $\mathrm{grad}: C^1(\mathbb{R}^n, \mathbb{R})\rightarrow C^0(\mathbb{R}^n, \mathbb{R}^n)$} &  $\mathrm{grad} G(x)=\Big( \frac{\partial G(x)}{\partial x_k}; 1\leq k \leq n\Big)$\\
		\hline
		\makecell[l]{basic rotation density operator\\ $\overline{\mathrm{ROT}}_{ij}: C^1(\mathbb{R}^n, \mathbb{R}^n)\rightarrow C^0(\mathbb{R}^n, \mathbb{R})$} & $\overline{\mathrm{ROT}}_{ij}f(x):=\frac{\partial f_i(x)}{\partial x_j}-\frac{\partial f_j(x)}{\partial x_i}$ \\
		\hline
		\makecell[l]{rotation density operator\\$\overline{\mathrm{ROT}}: C^1(\mathbb{R}^n, \mathbb{R}^n)\rightarrow C^0(\mathbb{R}^n, \mathbb{R}^{n^2})$}& $\overline{\mathrm{ROT}}f(x):=\Big[\overline{\mathrm{ROT}}_{ij}f(x)\Big]=\Big[\frac{\partial f_i(x)}{\partial x_j}-\frac{\partial f_j(x)}{\partial x_i}; 1\leq i,j\leq n\Big]$ \\
		\hline
		\makecell[l]{basic rotation density\\ $\rho_{i,j}\in C^1(\mathbb{R}^n, \mathbb{R})$}& $\rho_{ij}(x)=-\rho_{ji}(x)=\overline{\mathrm{ROT}}_{ij}f(x); 1\leq i,j\leq n$   \\
		\hline
		\makecell[l]{rotation density\\ $\rho\in C^1(\mathbb{R}^n, \mathbb{R}^{n^2})$}& $\rho(x)=\Big[\rho_{ij}(x); 1\leq i,j\leq n\Big]=\overline{\mathrm{ROT}}f(x)$   \\
		\hline
		\makecell[l]{basic rotation potentials\\ $R_{ij}\in C^3(\mathbb{R}^n, \mathbb{R})$}& $R_{ij}(x)=\overline{\int}\rho_{ij}(x); 1\leq i,j\leq n$ \\
		\hline
		\makecell[l]{rotation potential\\ $R\in C^3(\mathbb{R}^n, \mathbb{R}^{n^2})$}& $R(x)=\Big[R_{ij}(x); 1\leq i,j\leq n\Big]$ \\
		\hline
		\makecell[l]{basic rotation operator\\$\mathrm{ROT}_{ij}: C^1(\mathbb{R}^n, \mathbb{R})\rightarrow C^0(\mathbb{R}^n, \mathbb{R}^n)$} &  $\mathrm{ROT}_{ij} R_{ij}(x)=\Big( \delta_{ik}\frac{\partial R_{ij}(x)}{\partial x_j} - \delta_{jk}\frac{\partial R_{ij}(x)}{\partial x_i}; 1\leq k \leq n\Big)$ \\
		\hline
		\makecell[l]{rotation operator\\ $\mathrm{ROT}: C^1(\mathbb{R}^n, \mathbb{R}^{n^2})\rightarrow C^0(\mathbb{R}^n, \mathbb{R}^n)$ }&  $\mathrm{ROT} R:=\sum_{1\leq i,j\leq n}\frac{1}{2}\mathrm{ROT}_{ij} R_{ij}(x)=\Big( \sum_m\frac{\partial R_{km}(x)}{\partial x_m}; 1\leq k \leq n\Big)$\\
		\hline
	\end{tabular}
\end{table}

\newpage
\subsection{Supplementary Figure 1: schematic diagram of HHD}

\begin{figure}[!ht]
	\centering
	\includegraphics[width=0.9\linewidth]{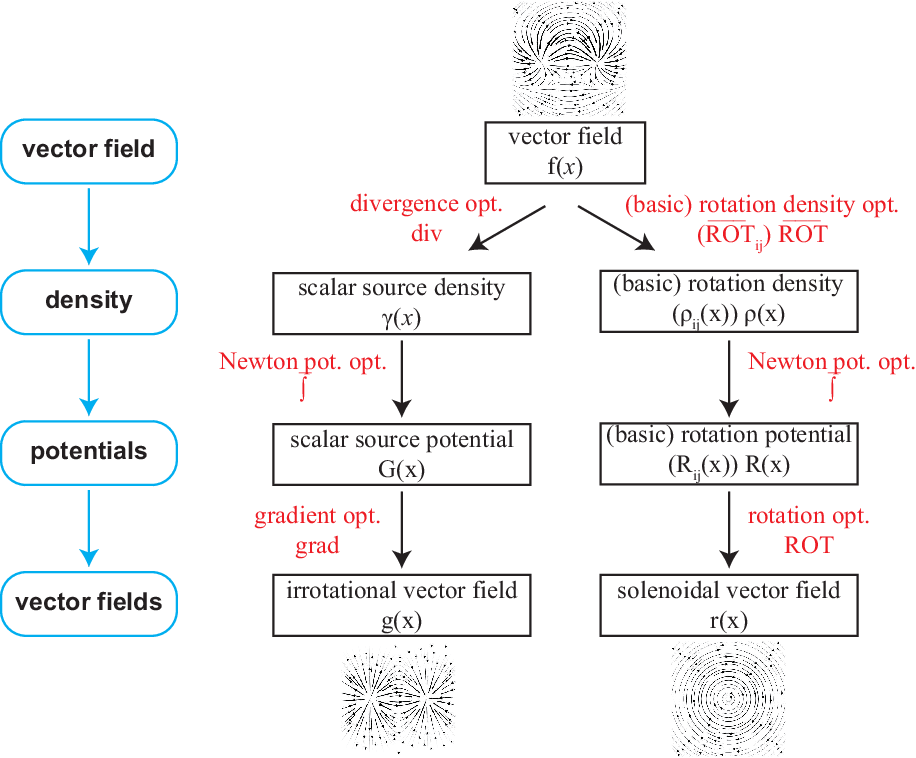}
	\caption{Schematic diagram of Helmholtz-Hodge Decomposition of vector field in $\mathbb{R}^n$.}
	\label{sfig-1}
\end{figure}

\end{document}